\documentclass[final,3p,times,twocolumn]{elsarticle}

\usepackage{ecrc}

\volume{00}

\firstpage{1}

\journalname{Nuclear Physics B Proceedings Supplement}

\runauth{}

\jid{nuphbp}

\jnltitlelogo{Nuclear Physics B Proceedings Supplement}

\usepackage{amssymb}

\usepackage{lineno}

\usepackage[figuresright]{rotating}
\usepackage{fancyhdr}
\usepackage{underscore}
\catcode`\_=12\def\fileunderscore{_}\catcode`\_=8

\begin{document}

\begin{frontmatter}
\pagestyle{plain}
\fancyhead{}


\dochead{}

\title{Status and prospects for BSM ( (N)MSSM) Higgs searches at the LHC}


\author{M.P. Casado$^1$ on behalf of the ATLAS and CMS collaborations.}

\address{$^1$Departamet de F\'isica, Universitat Aut\`onoma de Barcelona, Institut de F\'isica d'Altes Energies and Barcelona Institute of Science and Technology, Spain.}
\address{\large \rm Talk presented at the International Workshop on Future Linear Colliders (LCWS15), Whistler, Canada, 2-6 November 2015}
\begin{abstract}
Searches for Beyond the Standard Model Higgs processes in the context of Minimal Supersymmetric Standard Model and Next to MSSM are presented.
The results are based on the first LHC run of pp collision data recorded by the ATLAS and CMS experiments at the CERN Large Hadron Collider at 
centre-of-mass energies of 7 and 8 TeV, corresponding to integrated luminosities of about 5 and 20 fb$^{-1}$ respectively.  
Current searches constrain large parts of the parameter space. No evidence for BSM Higgs is found. 
\end{abstract}


\end{frontmatter}


\section{Introduction}
\label{introduction}
The discovery of a new particle~\cite{ATLAS,CMS} in the search for the Standard Model (SM)~\cite{RefSM} Higgs boson at the Large Hadron Collider (LHC)~\cite{LHC}, performed with the ATLAS~\cite{ATLASDet} and CMS~\cite{CMSDet} detectors in July 2012, was an important achievement in the
un\-der\-stan\-ding of the electroweak symmetry breaking me\-cha\-nism~\cite{EWSymBreak}.

With this boson the Standard Model (SM) of particle physics is complete. However many effects remain unexplained in the SM, as the
hierarchy problem or the existence of Dark Matter.

\begin{figure}[!htpb]
  \begin{center}
    \includegraphics[width=0.40\textwidth]{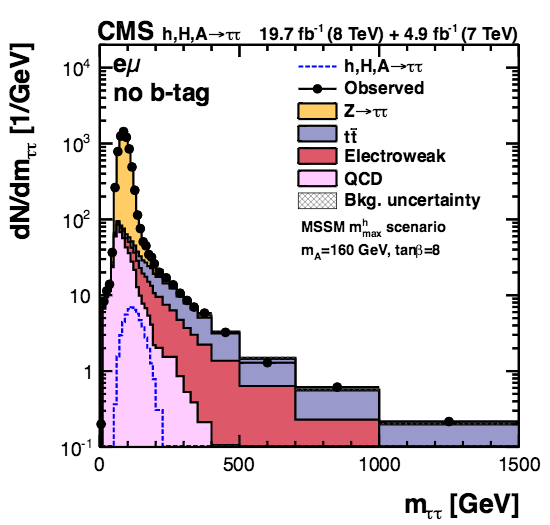}
  \end{center}
  \caption{
Invariant mass of the $\tau\tau$ pair in the electron, muon channel, no b-tag category in the $h/H/A \rightarrow \tau\tau$ analysis~\cite{CMSNeutralHiggs}. The main backgrounds are $Z\rightarrow \tau\tau$ processes and QCD events. There is good agreement between the observed data and the prediction.}
\label{HToTauTauInvMassCMS}
\end{figure}
\begin{figure*}[!htbp]
  \begin{center}
    \includegraphics[width=0.80\textwidth]{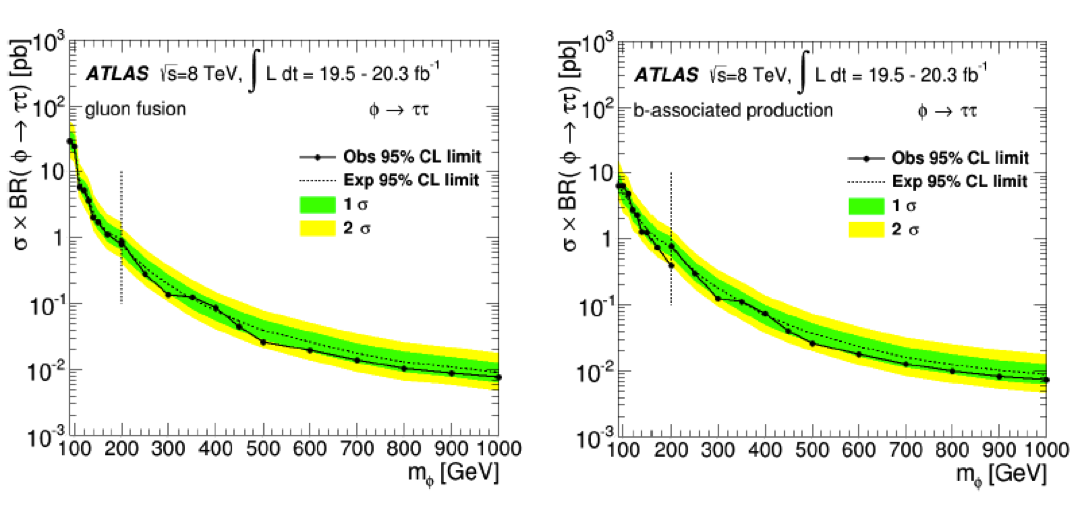}
  \end{center}
  \caption{
Observed and expected 95\% CL limits on the cross-section times de branching ratio to a $\tau\tau$ pair as a function of the Higgs boson mass (gluon-fusion production on the left, b-associated production on the right) for the $h/H/A \rightarrow \tau\tau$ analysis~\cite{ATLASNeutralHiggs}.  No excess of events is found.}
\label{HToTauTauResultATLAS}
\end{figure*}
Moreover, there is no theoretical reason to restrict the model to only one Higgs boson: (1) the generation of fermion masses could also be realized by more bosons~\cite{OnlyOneHiggs}, (2) many theories include extra Higgs boson(s), as Supersymmetry (SUSY), models with axions, baryogenesis, neutrino masses, etc. 

In this paper we will concentrate on Higgs bosons within the context of Minimal Supersymmetric Standard Model (MSSM)~\cite{HplusModels}  and Next to Minimal Supersymmetric Standard Model (NMSSM)~\cite{RefNMSSM}.

\section{Searches within the Minimal Supersymmetric Standard Model}
\label{MSSM}

\subsection{MSSM Neutral Higgs bosons at LHC}

This search is important at high $\tan \beta$ (ratio of vacuum expectation values of the two Higgs doubles in the MSSM).
The neutral Higgs field can be produced via gluon-fusion and in association with a b quark. The two most relevant final states are  $h/H/A \rightarrow \tau\tau$ and bb. The $\tau\tau$ modes tend to be more sensitive due to the better performance of the tau-jet selection.

In the $h/H/A \rightarrow \tau\tau$ search a categorization is applied depending on the event properties ($\tau\tau$ decay and 
``b-tag"/``b-veto"). ATLAS performed the search using 8~TeV run 1 data~\cite{ATLASNeutralHiggs} while
CMS combined 7 and 8~TeV datasets~\cite{CMSNeutralHiggs}.

Figure~\ref{HToTauTauInvMassCMS} presents the invariant mass of the $\tau\tau$ pair for the case in which one $\tau$ decays to electron and
the other decays to a muon (``electron-muon" channel) and for the b-veto case in the CMS analysis.
The main backgrounds are $Z\rightarrow \tau\tau$ processes and QCD events. There is good agreement between the observed data and the prediction. No excess of events is found.

Figure~\ref{HToTauTauResultATLAS} shows the observed and expected 95\% CL limits on the cross-section times branching ratio to a $\tau\tau$ pair as a function of the Higgs boson mass as obtained by ATLAS.

\subsection{ATLAS and CMS search for H$^{\pm}$}
This search covers the whole spectrum of $\tan\beta$. 
The $H^{\pm}$ can decay to $\tau\nu$, tb, cs, hW, etc. with different branching ratios depending on $\tan\beta$.
The decay to $H^{\pm} \rightarrow \tau\nu$ is relevant in a large parameter range, specially for low $m_{H^{\pm}}$ (below $m_{top}$). For $m_{H^{\pm}}$ above $m_{top}$ $H^{\pm} \rightarrow tb$ is the predominant decay.

\begin{figure}[!htpb]
  \begin{center}
    \includegraphics[width=0.40\textwidth]{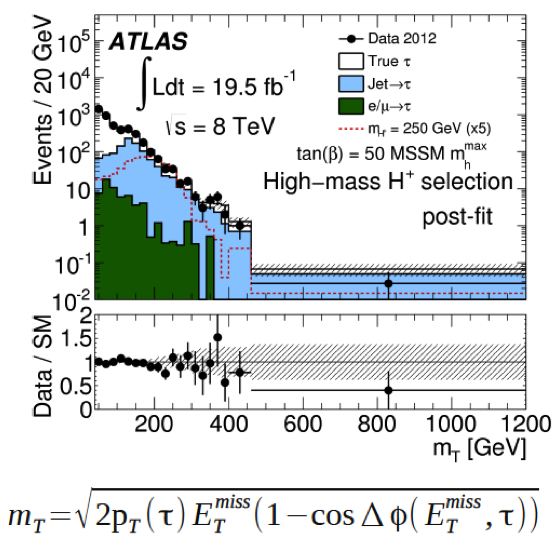}
  \end{center}
  \caption{
Transverse invariant mass between visible $\tau$ decay products and the missing $E_T$ of the event for the search of $H^{\pm} \rightarrow \tau\nu$~\cite{ATLASHpHiggs}.
}
\label{HplusInvMassATLAS}
\end{figure}

\begin{figure}[!htpb]
  \begin{center}
    \includegraphics[width=0.40\textwidth]{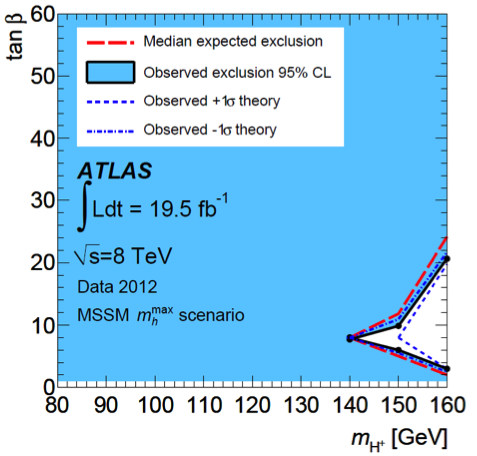}
  \end{center}
  \caption{
95\%CL exclusion limits on $\tan\beta$ as a function of $m_H^{\pm}$ is shown for $m_H^{\pm} < m_{top}$ in the $H^{\pm} \rightarrow \tau\nu$ search~\cite{ATLASHpHiggs}.
The interpretation is performed in the context of the $m_h^{max}$ benchmark scenario~\cite{HplusModels} of the MSSM.
}
\label{HplusResultATLAS}
\end{figure}

ATLAS and CMS perform this search~\cite{ATLASHpHiggs},~\cite{CMSHpHiggs} u\-sing the full 8~TeV Run 1 sample. The strategies are similar in both analyses. High and low mass $H^{\pm}$ categories are treated separately and a tau + missing $E_T$ trigger is used to trigger events. The most important discrimina\-ting variable is presented in Fig.~\ref{HplusInvMassATLAS} for the ATLAS search. The 95\%~CL exclusion limits on $\tan\beta$ as a function of $m_H^{\pm}$ is shown for $m_H^{\pm} < m_{top}$ in Fig.\ref{HplusResultATLAS}.	


\begin{figure}[!htpb]
  \begin{center}
    \includegraphics[width=0.40\textwidth]{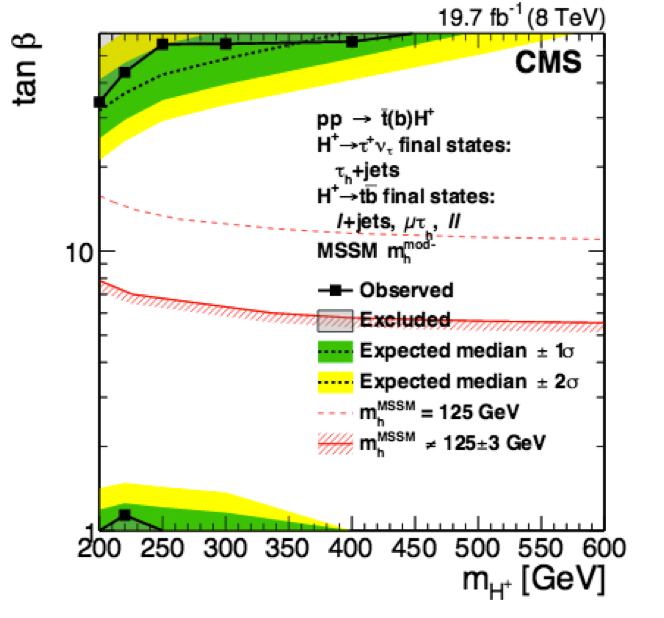}
  \end{center}
  \caption{
Observed and excluded limits on the $\tan\beta$-$m_H^{\pm}$ plane in the $H^{\pm} \rightarrow \tau\nu$ search~\cite{CMSHpHiggs} using the 
$m_h^{mod-}$ benchmark scenario~\cite{HplusModels} of the MSSM.
. The excluded area by the discovery of the SM 125-GeV Higgs boson is drawn as the red line.}
\label{HplusResultCMS}
\end{figure}

CMS has combined $H^{\pm} \rightarrow \tau\nu$ and $H^{\pm} \rightarrow tb$ channels. The observed and excluded limits on the $\tan\beta$-$m_{H^{\pm}}$ plane are presented in Fig.\ref{HplusResultCMS}. In this plot the excluded area by the discovery of the SM 125-GeV Higgs boson is drawn as a dotted red line
and as a continuous red line if an error of $\pm 3 GeV$ is included.

CMS has a recent analysis~\cite{CMSHcsHiggs} searching for $m_H^{\pm}\rightarrow cs$. No excess of events is found.

\subsection{Search for $A\rightarrow Zh$ at the LHC}
This search is important at low $\tan \beta$ and has been performed both by ATLAS~\cite{ATLASAZhHiggs} and CMS~\cite{CMSAZhHiggs}. 
Possible final states are
$ll\tau\tau / llbb / \nu\nu bb$.
It takes advantage of the $Z\rightarrow ll / Z \rightarrow \nu\nu / h$ resonances and ratios of the Higgs boson decays ($b\bar{b} /\tau\tau$).

\begin{figure}[!htpb]
  \begin{center}
    \includegraphics[width=0.40\textwidth]{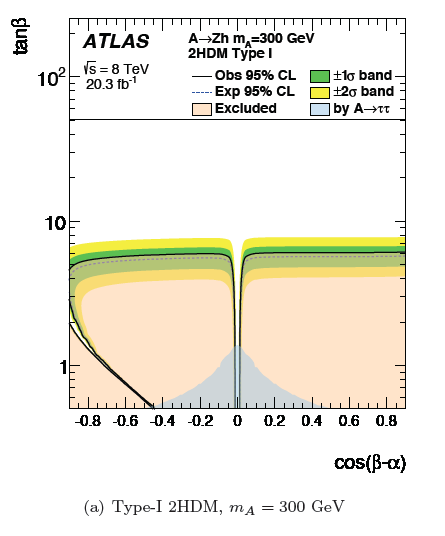} 
     \includegraphics[width=0.40\textwidth]{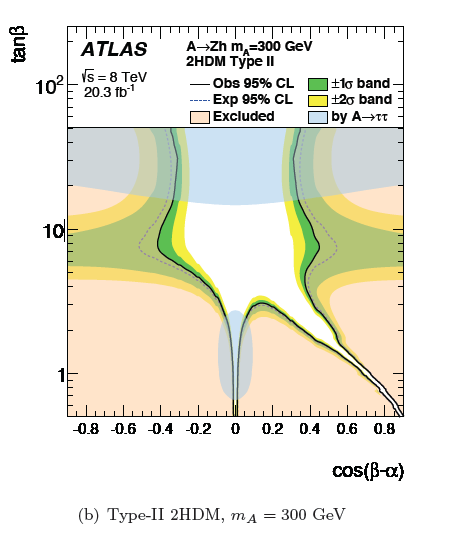}
 \end{center}
  \caption{
Observed and expected 95\% CL limits in the $\tan\beta$ - $\cos (\beta - \alpha)$ plane for two MSSM scenarios~\cite{HplusModels} (Type I and Type II) in the $A\rightarrow Zh$ search~\cite{ATLASAZhHiggs}. The angle $\alpha$ is the CP-even Higgs mixing angle in the MSSM.}
\label{AZhResultATLAS}
\end{figure}

Figure~\ref{AZhResultATLAS} shows the observed and expected 95\% CL limits in the $\tan\beta$ - $\cos (\beta - \alpha)$ plane for two MSSM scenarios (Type I and Type II). The angle $\alpha$ is the CP-even Higgs mixing angle in the MSSM. No excess of events is observed.

\subsection{Search for $hh$ processes}
This search is done by ATLAS~\cite{ATLAShhHiggs} and CMS~\cite{CMShhHiggs} in resonant and non-resonant Higgs boson pair production. The considered final states are $bb\gamma\gamma / bbbb / bb\tau\tau / WW \gamma\gamma$.

\begin{figure}[!htpb]
  \begin{center}
    \includegraphics[width=0.40\textwidth]{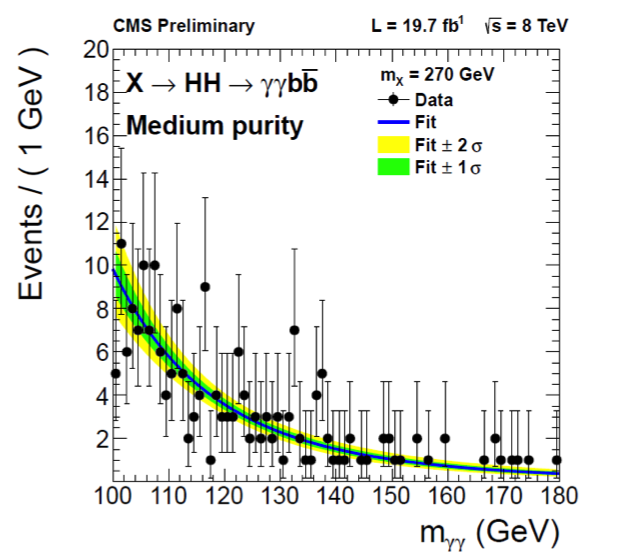}
  \end{center}
  \caption{
Non-resonant background fits in the $m_{\gamma\gamma}$ for one of the
categories for the resonance mass hypothesis of 270~GeV in the $hh$ CMS analysis~\cite{CMShhHiggs}.}
\label{hhInvMassCMS}
\end{figure}

Figure~\ref{hhInvMassCMS} shows the non-resonant background fit in the 
$m_{\gamma\gamma}$ for one of the
final states for the resonance mass hypothesis of 270~GeV in the CMS analysis.

\subsection{Search for $H\rightarrow WW / ZZ$ processes}
This search is done by ATLAS~\cite{ATLASDibosonHiggs} and CMS~\cite{CMSDibosonHiggs}. 

\begin{figure}[!htpb]
  \begin{center}
    \includegraphics[width=0.40\textwidth]{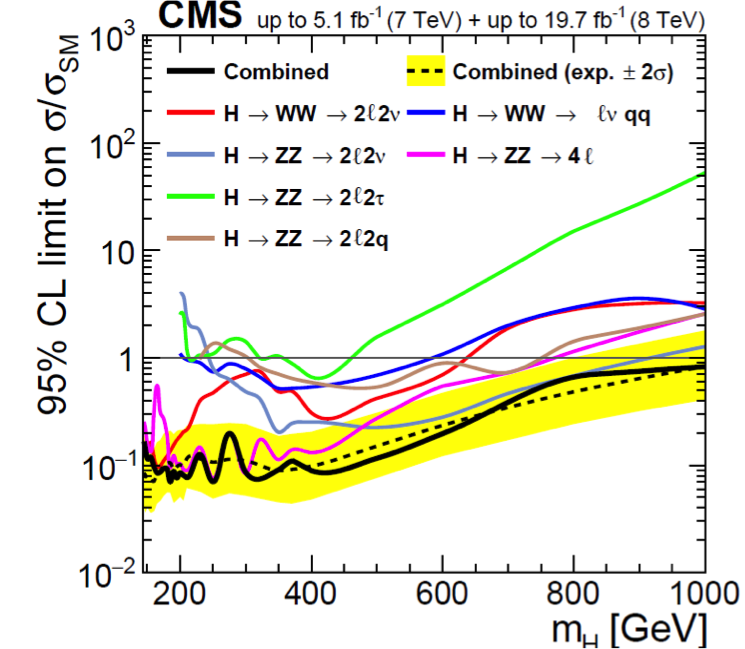}
  \end{center}
  \caption{
 95\% CL on the ratio of cross-section over the SM cross-section for each of the contributing final states
in the $H\rightarrow WW / ZZ$ search~\cite{CMSDibosonHiggs}. The theoretical cross section,  $\sigma_{SM}$, is computed in~\cite{1307.1347}.
}
\label{DibosonResultCMS}
\end{figure}

The invariant masses obtained in the analysis allow to set upper 95\% CL limits on the production cross-section. The different results from the different final states and the combination are presented in Fig.~\ref{DibosonResultCMS} for CMS.
\section{Searches within the Next to Minimal Supersymmetric Standard Model}
\subsection{Search for $a\rightarrow \mu\mu$}
\begin{figure}[!htpb]
  \begin{center}
    \includegraphics[width=0.40\textwidth]{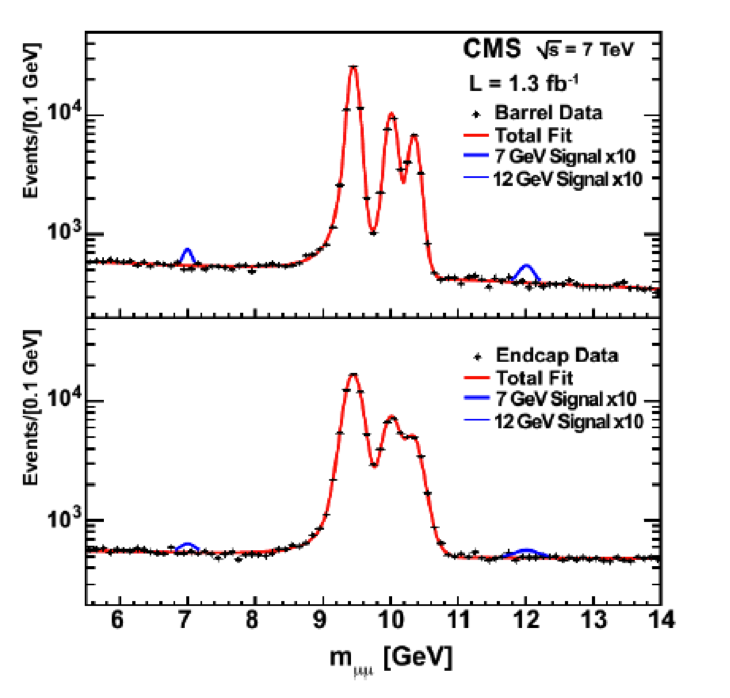}
  \end{center}
  \caption{
Invariant mass of the dimuon pair for barrel (up) and endcap (bottom) in the $a\rightarrow \mu\mu$ search~\cite{CMSamumuHiggs}. The effect of a possible signal at 7 and 12~GeV is shown with the blue line.}
\label{amumuInvMassCMS}
\end{figure}

This is a low mass search performed by CMS in~\cite{CMSamumuHiggs}. The $a$ boson is produced in gluon-fusion and searched for in the decay $\mu\mu$. In Fig.~\ref{amumuInvMassCMS} the invariant mass of the dimuon pair is shown for barrel and endcap. The effect of a possible signal at 7 and 12~GeV is shown with the blue line.
No excess of events is found in this search.

\subsection{Search for $h\rightarrow aa \rightarrow \mu\mu\tau\tau / \mu\mu\mu\mu$}
\begin{figure}[!htpb]
  \begin{center}
    \includegraphics[width=0.40\textwidth]{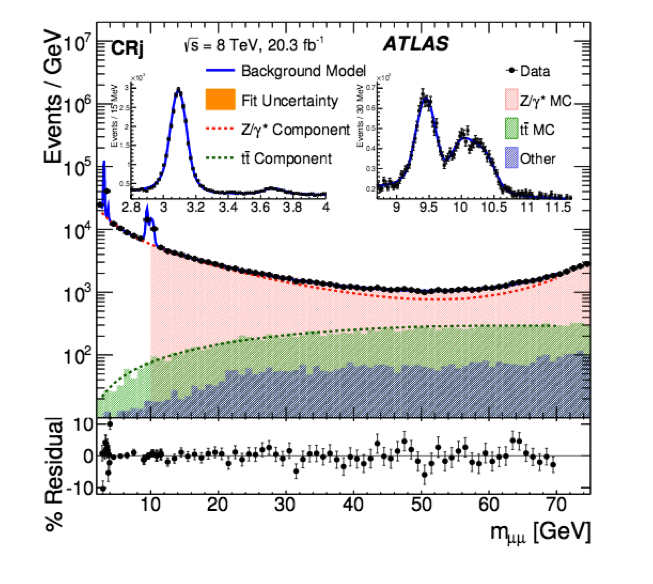}
  \end{center}
  \caption{
Invariant mass of the dimuon pair for data and MC in the $h\rightarrow aa \rightarrow \mu\mu\tau\tau / \mu\mu\mu\mu$ search~\cite{ATLAShaaHiggs}. On the top part of the Figure the light resonances are depicted showing the good understanding of the data.}
\label{hToaaInvMassATLAS}
\end{figure}

This search is performed by ATLAS~\cite{ATLAShaaHiggs} and CMS in~\cite{CMShaaHiggs}. The final states can be $\mu\mu\tau\tau / \mu\mu\mu\mu$.  Fig.~\ref{hToaaInvMassATLAS} shows the invariant mass of the dimuon pair for data and MC. On the top part of the Figure the light resonances are depicted showing the good understanding of the data. 

\section{Conclusions}
Current searches constrain large parts of the parameter space. So far there is no evidence for Beyond the Standard Model Higgs. However, there are still many possibilities to explore and many searches are still starting up, and this will be a relevant area in Run 2.
\nocite{*}
\bibliographystyle{elsarticle-num}
\bibliography{martin}



\end{document}